\documentclass[12pt]{iopart}
\usepackage{iopams}
\usepackage{epsfig,graphicx}
\begin{document}

\title{Noise driven translocation of short polymers in crowded solutions}

\author{N. Pizzolato$^a$, A. Fiasconaro$^{ab}$\footnote{Email: afiasconaro@gip.dft.unipa.it},
B. Spagnolo$^a$\footnote{Email: spagnolo@unipa.it}}
\address{$^a$Dipartimento di Fisica e Tecnologie Relative, \\
Universit\`a di Palermo and CNISM-INFM, Group of
Interdisciplinary Physics\footnote{URL: http://gip.dft.unipa.it}, \\
Viale delle Scienze, edificio 18, I-90128 Palermo, Italy\\
$^b$Mark Kac Complex Systems Research Center, Institute of
Physics,\\
Jagellonian University, Reymonta 4, 30-059 Krak$\rm\acute{o}$w,
Poland} \ead{npizzolato@gip.dft.unipa.it}

\begin{abstract}
In this work we study the noise induced effects on the dynamics of
short polymers crossing a potential barrier, in the presence of a
metastable state. An improved version of the Rouse model for a
flexible polymer has been adopted to mimic the molecular dynamics by
taking into account both the interactions between adjacent monomers
and introducing a Lennard-Jones potential between all beads. A
bending recoil torque has also been included in our model. The
polymer dynamics is simulated in a two-dimensional domain by
numerically solving the Langevin equations of motion with a Gaussian
uncorrelated noise. We find a nonmonotonic behaviour of the mean
first passage time and the most probable translocation time, of the
polymer centre of inertia, as a function of the polymer length at
low noise intensity. We show how thermal fluctuations influence the
motion of short polymers, by inducing two different regimes of
translocation in the molecule transport dynamics. In this context,
the role played by the length of the molecule in the translocation
time is investigated.
\end{abstract}

\maketitle

\section{Introduction}\label{sect1}
The knowledge of the translocation dynamics of a molecule moving
across a membrane or surmounting a barrier is a fundamental step
towards the full comprehensions of the functioning of many
biological systems. In the cell environment, DNA and RNA translocate
across nuclear pores and many proteins work on the bases of their
ability to go beyond a potential barrier. Reciprocal translocation
between portions of chromosomes causes the genetic mutations which
characterize several tumors \cite{Leng1998, Sover2006}. The cancer
target therapy is based on a drug delivery mechanism that crucially
depends on the translocation time of the chemotherapeutic molecules
\cite{Tseng2002, Garcia2005, Yotsu2008}. Polymer translocation has
been found to play a key role in the acquisition of multi-drug
resistance to cancer therapy \cite{Higgins2007}. The study of the
transport of macromolecules across a nanopore is also important for
technological applications. For example, the development of
chemomechanical actuators \cite{Sunda2008}, the separation of
molecules by liquid chromatography \cite{Peyrin2001, Keller2007} or
fast DNA sequencing techniques \cite{Mannion2006} are both performed
by forcing the molecule to move inside nano-channel devices.

Many experimental studies on polymer translocation have been carried
out in the wake of the pioneering work of Kasianowicz and
collaborators (1996). In these experiments, single-stranded DNA
(ssDNA) molecules are forced by a voltage bias to pass through an
$\alpha$-hemolysin ($\alpha$-HL) pore imbedded into a membrane. The
passage of the molecule inside the protein channel causes a
reduction of the electrolyte ion current and the duration time of
any current blockade is recorded. In this way, a linear relationship
of the most probable crossing time $\tau_{\rm{p}}$ with the molecule
length and an inverse proportionality law between $\tau_{\rm{p}}$
and the applied voltage were established \cite{Kasianowicz1996}. The
$\alpha$-HL channel device has been extensively used for probing the
transport dynamics of several types of polynucleotide (DNA or RNA)
molecules having same length but different adenine and cytosine
contents \cite{Akeson1999, Meller2000, Meller2002}. A different
nucleotide composition of the DNA brings about a significant change
in the distributions of the duration times of the corresponding
current drops. These findings confirm that the dynamics of
biopolymer translocation across an $\alpha$-HL channel is governed
by pore-molecule interactions, which are dependent on the details of
DNA sequences \cite{Luo2008}, the orientation and the driving
voltage \cite{Wanunu2008}. Moreover, $\tau_{\rm p}$ scales as the
inverse square of the temperature and the differences on the
crossing time due to different nucleotide compositions are
progressively reduced when the temperature increases
\cite{Meller2000}.

Recent advances in semiconductor technology have enabled the
construction of synthetic nanopores for DNA translocation
experiments \cite{Li2001, Li2003}. By using a silicon-oxide
nanopore, it was found a power-law relationship between $\tau_{\rm
p}$ and the polymer length, with an exponent of 1.27
\cite{Storm2005a, Storm2005b}. However, a retarding effect on the
crossing time has been observed as an effect of the voltage dragging
itself, which could be stronger enough to disturb the biopolymer on
finding the best orientation for the translocation \cite{Aksi2004}.
In fact, other experiments show that DNA molecule's translocation
speed in a solid-state nanopore is strongly dependent on electrolyte
temperature, salt concentration, viscosity and applied voltage bias
\cite{Fan2005, Folo2005}.

In contrast to the proportionality law between the translocation
time and the polymer length, experiments on the transport dynamics
of DNA molecules driven inside an entropic trap array have shown
longer crossing times for shorter molecules, suggesting the
existence of a quasi-equilibrium state of the polymer during the
passage \cite{Han1999,Han2000}.

The complex scenario of the translocation dynamics coming from
experiments has been enriched by several theoretical and simulation
studies \cite{Sung1996, Muthu1999, Lubensky1999, Tian2003, Aksi2004,
Luo2006, Huop2007, Luo2007, Luo2008}. A power law relationship
between the translocation time of a long $N$-segments chain molecule
and the polymer length has been found, with an exponent of 2 in the
presence of a free energy bias between the two sides of the membrane
or 3 in the lack of adsorption \cite{Park1998}. The relationship
between the crossing time and $N$ becomes linear in the presence of
strong interactions between the polymer and the pore walls
\cite{Lubensky1999} or in the case of an hairpin crossing mechanism
of translocation \cite{Sebastian2006}.

In spite of the above contributions, the complicated biological
environments and boundary conditions make the problem of polymer
translocation still far from a common understanding and, in this
framework, a detailed description of the transport dynamics of short
chain molecules is missing. The motion of a polymer passing through
a pore takes place in solutions where thermal fluctuations always
affect the translocation dynamics. In this paper we present the
results of our studies on the noise-induced effects on the transport
dynamics of short polymers surmounting a potential barrier, in the
presence of a metastable state. Molecular dynamics simulations are
performed by modeling the polymer as a flexible chain molecule with
harmonic interactions between adjacent monomers and a Lennard-Jones
(LJ) potential between all beads. A bending recoil torque has also
been included in our model. The polymer dynamics is simulated in a
two-dimensional domain by numerically solving the Langevin equations
of motion with a Gaussian uncorrelated noise. The dependence of the
mean first passage time (MFPT) of the polymer centre of inertia from
the length $L$ of the chain molecule critically changes with the
noise intensity \cite{Pizz2008}. In this paper we focus our research
on short polymers, which usually show two different regimes of
translocation depending on the ratio of $L$ over the length of the
channel \cite{Meller2001}. We find a non-monotonic behaviour of both
MFPT and $\tau_{\rm p}$ as a function of polymer length at low noise
intensity. In this context, the role played by the stiffness of the
molecule in the translocation dynamics is also investigated. In
section \ref{sect2} we present our polymer chain model and give the
details of the molecular dynamics simulations. The final results are
described in section \ref{sect3} and discussed in section
\ref{sect4}.

\section{Model and method}\label{sect2}
\subsection{The polymer chain model}\label{subsect21}
In our simulations, the polymer is modeled by a semi-flexible chain
of $N$ beads connected by harmonic springs \cite{Rouse1953}. The
contour length is defined as $L_0=Nd$, where $d$ is the equilibrium
distance between adjacent monomers. Both excluded volume effect and
van der Waals interactions between all beads are kept into account
by introducing a Lennard-Jones potential. In this work, we
investigate the dynamics of a linear chain molecule. In order to
confer a suitable stiffness to the chain, a bending recoil torque is
included in the model, with a rest angle $\theta_0=0$ between two
consecutive bonds. The polymer motion in the liquid solvent induces
a velocity field which is felt by all the beads. To first order, we
neglect this hydrodynamic effect. Such approximation implies that
our model cannot be used to mimic the molecule behaviour in dilute
polymeric solutions, but it appears to be much more appropriate for
polymeric melts \cite{Tothova2005}. The potential energy of the
modeled chain molecule is
\begin{eqnarray}
U=U_{\rm Har}+U_{\rm Bend}+U_{\rm LJ} \label{eq1}
\end{eqnarray}
where $U_{\rm Har}$ represents the energy required to extend the
bond between two consecutive beads, $U_{\rm Bend}$ the work required
to bend the chain, and $U_{\rm LJ}$ the Lennard-Jones potential.
Respectively, we have
\begin{eqnarray}
U_{\rm Har}=\sum_{i=1}^{N-1} K_{\rm r}(r_{i,i+1}-d)^2 \label{eq2} \\
U_{\rm Bend}=\sum_{i=2}^{N-1} K_{\rm \theta}(\theta_{i-1,i+1}-\theta_0)^2 \label{eq3} \\
U_{\rm LJ}=4\epsilon_{\rm LJ}\sum_{i,j (i\neq
j)}\left[\left(\frac{\sigma}{r_{ij}}\right)^{12}-\left(\frac{\sigma}{r_{ij}}\right)^6\right]
\label{eq4}
\end{eqnarray}
where $K_{\rm r}$ is the elastic constant, $r_{ij}$ the distance
between particles $i$ and $j$, $K_{\rm \theta}$ the bending modulus,
$\epsilon_{\rm LJ}$ the LJ energy depth and $\sigma$ the monomer
diameter.

\subsection{Molecular dynamics simulations}\label{subsect22}
\begin{figure}[htbp]
\includegraphics[width=10cm,height=10cm]{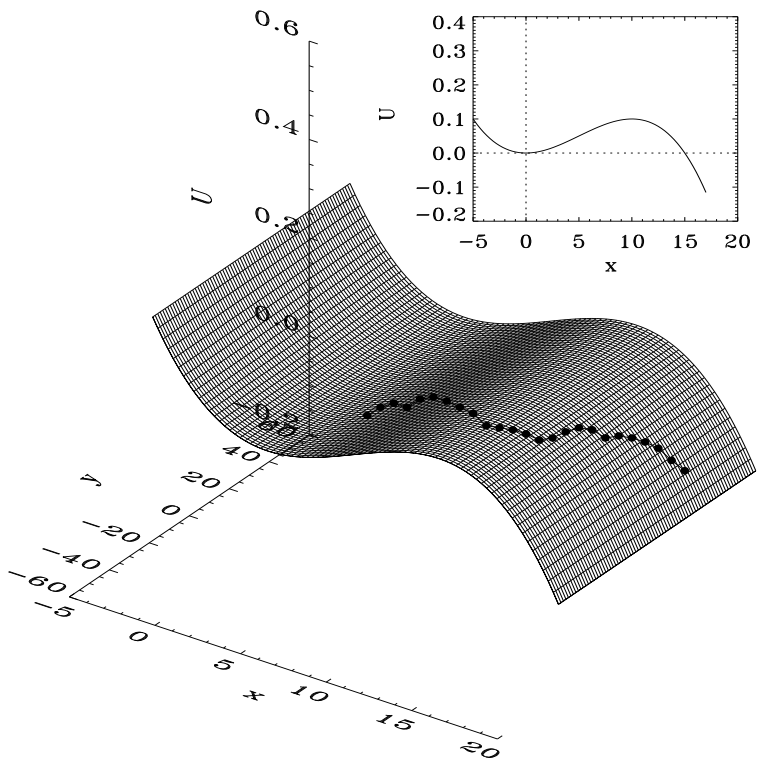}
\caption{\small 3D-view and projection on z-x plane of the potential
energy $U_{\rm Ext}$, which is included in our system to simulate
the presence of a barrier to be surmounted by the polymer. A sketch
of the translocating chain molecule is shown.} \label{fig1}
\end{figure}

The effect of temperature fluctuations on the dynamics of a chain
polymer escaping from a metastable state is studied in a
two-dimensional domain. The polymer motion is modeled as a
stochastic process of diffusion in the presence of a potential
barrier having the form:
\begin{eqnarray}
U_{\rm Ext}(x)=ax^2-bx^3 \label{eq5}
\end{eqnarray}
with parameters $a=3\cdot10^{-3}$ and $b=2\cdot10^{-4}$. A
three-dimensional view of $U_{\rm Ext}$ is plotted in figure
\ref{fig1}. The drift of the $i^{\rm th}$ monomer of the chain
molecule is described by the following overdamped Langevin
equations:
\begin{eqnarray}
\gamma\frac{dx}{dt}&=&-\frac{\partial{U}}{\partial{x}}-\frac{\partial{U_{\rm
Ext}(x)}}{\partial{x}}+\sqrt{D}\xi_{\rm x}(t)
\label{eq6}\\
\gamma\frac{dy}{dt}&=&-\frac{\partial{U}}{\partial{y}}+\sqrt{D}\xi_{\rm
y}(t) \label{eq7}
\end{eqnarray}
where $U$ is the interaction potential, defined by equation
(\ref{eq1}), $\xi_x(t)$ and $\xi_y(t)$ are white Gaussian noise
modeling the temperature fluctuations, with the usual statistical
properties, namely $\langle\xi_k(t)\rangle=0$ and
$\langle\xi_k(t)\xi_l(t+\tau)\rangle=D\delta_{(k,l)}\delta(\tau)$
for $(k,l=x,y)$ and $\gamma$ is the friction coefficient. In this
work we set the $\gamma$ parameter equal to one. The standard
Lennard-Jones time scale is $\tau_{\rm LJ}=(m\sigma^2/\epsilon_{\rm
LJ})^{1/2}$, where $m$ is the mass of the monomer. A bead of a
single-stranded DNA is formed approximately by three nucleotide
bases and then $\sigma\sim 1.5$ nm and $m\approx 936$ amu
\cite{Luo2008}. Orders of magnitude of the quantities here involved
are nanometers for the characteristic lengths of the system (polymer
and barrier extension) and microseconds for the time domain.
Moreover, our simulation time $t_s$ is scaled with the friction
parameter as $t_s=t/\gamma$, therefore we use arbitrary units for
all computed translocation times. We set both $K_{\rm r}$ and
$K_{\rm \theta}$ equal to 10, $\epsilon_{\rm LJ}$ equal to 0.1 and
$\sigma$ equal to 3, in arbitrary units. The inter-beads rest length
$d$ is chosen equal to 5, while the number of monomers $N$ ranges
from 4 to 40 units.

A sequence of $10^3$ numerical simulations has been performed for
each of the 7 different values of the noise intensity $D$ and 11
different polymer lengths. The starting condition for all performed
molecular dynamics simulations provides that the initial distance
between two adjacent monomers of the chain is equal to the rest
length $d$ of the ideal spring connecting them. We have carried out
our study by selecting an initial spatial distribution of the
polymer with all monomers at the same $x$ coordinate equal to
$x_0$=0, which corresponds to the local minimum (metastable state)
of the potential energy of the barrier. Every simulation stops when
the $x$ coordinate of the center of mass of the chain reaches the
final position at $x_f$=15. This value has been chosen after
performing a set of simulations by assuming $x_f$=30 and the noise
intensities $D=0.3$, 1.0, 4.0 and 10.0, respectively. We have found
that the polymer center of mass takes a very short time to travel
from $x$=15 to $x_f$=30 in comparison with the average escape time
from $x_i$=0 to $x_f$=15. Therefore, by assuming $x_f$=15, we are
confident that the probability for the polymer to be trapped back
into the potential well is extremely low. The distribution of the
translocation times is analyzed and the MFPT, $\tau_{\rm p}$ and the
median of the distribution (hereafter 'the median') are calculated.

\section{Results}\label{sect3}
\begin{figure}[htbp]
\hspace{-0.3cm}
\includegraphics[width=5.42cm,height=4.5cm]{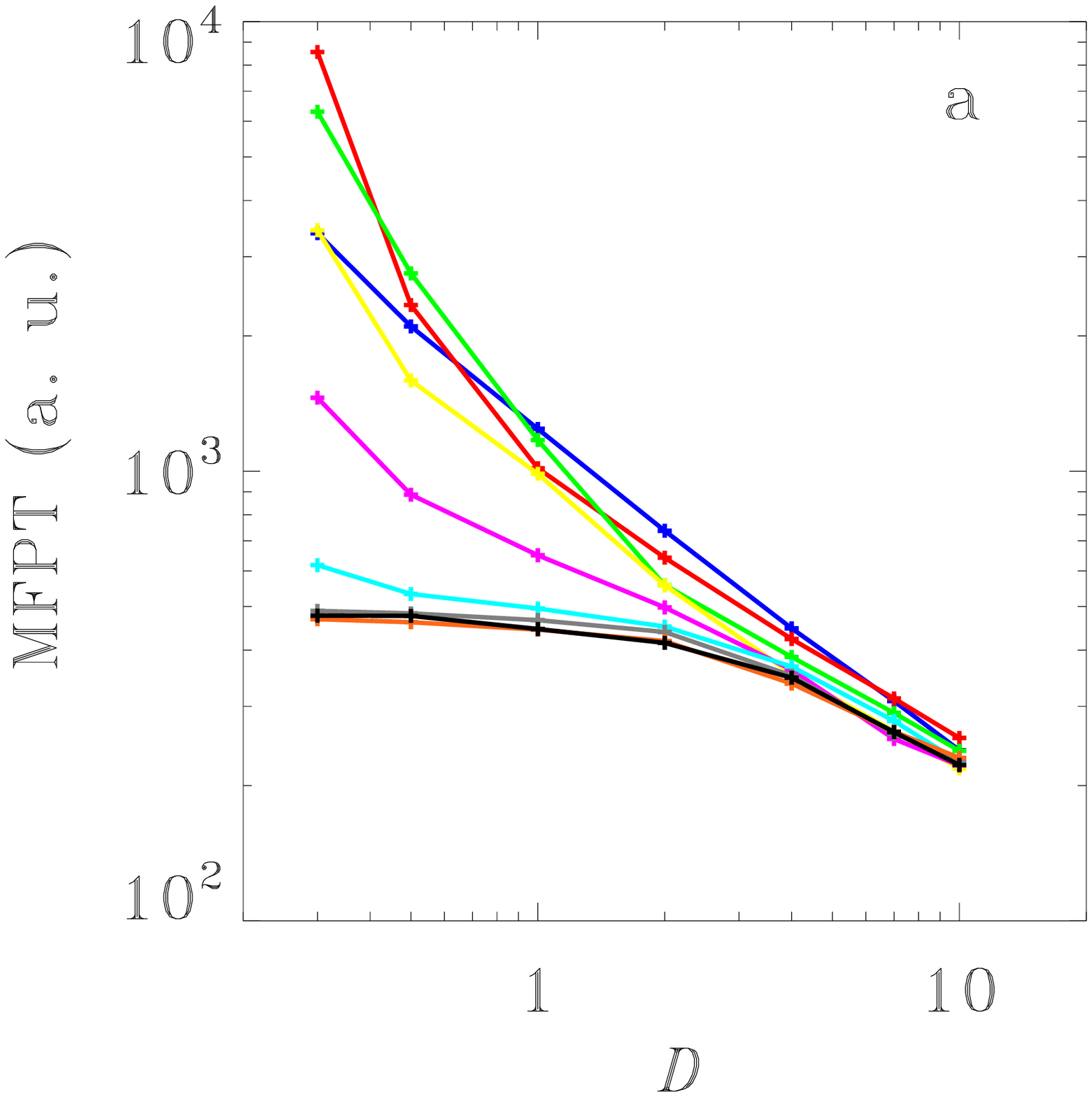}
\hspace{-0.38cm}
\includegraphics[width=5.42cm,height=4.5cm]{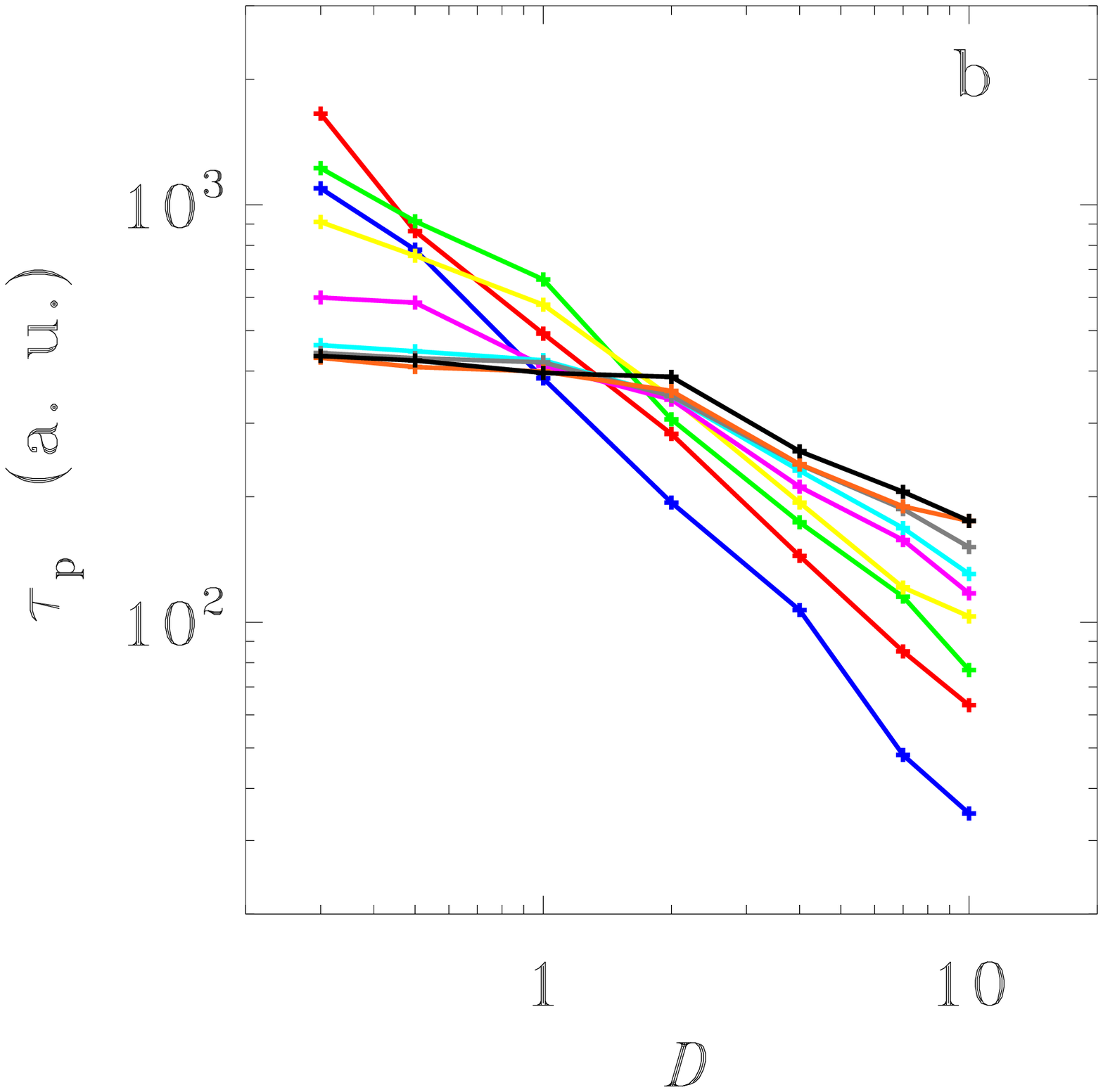}
\hspace{-0.38cm}
\includegraphics[width=5.42cm,height=4.5cm]{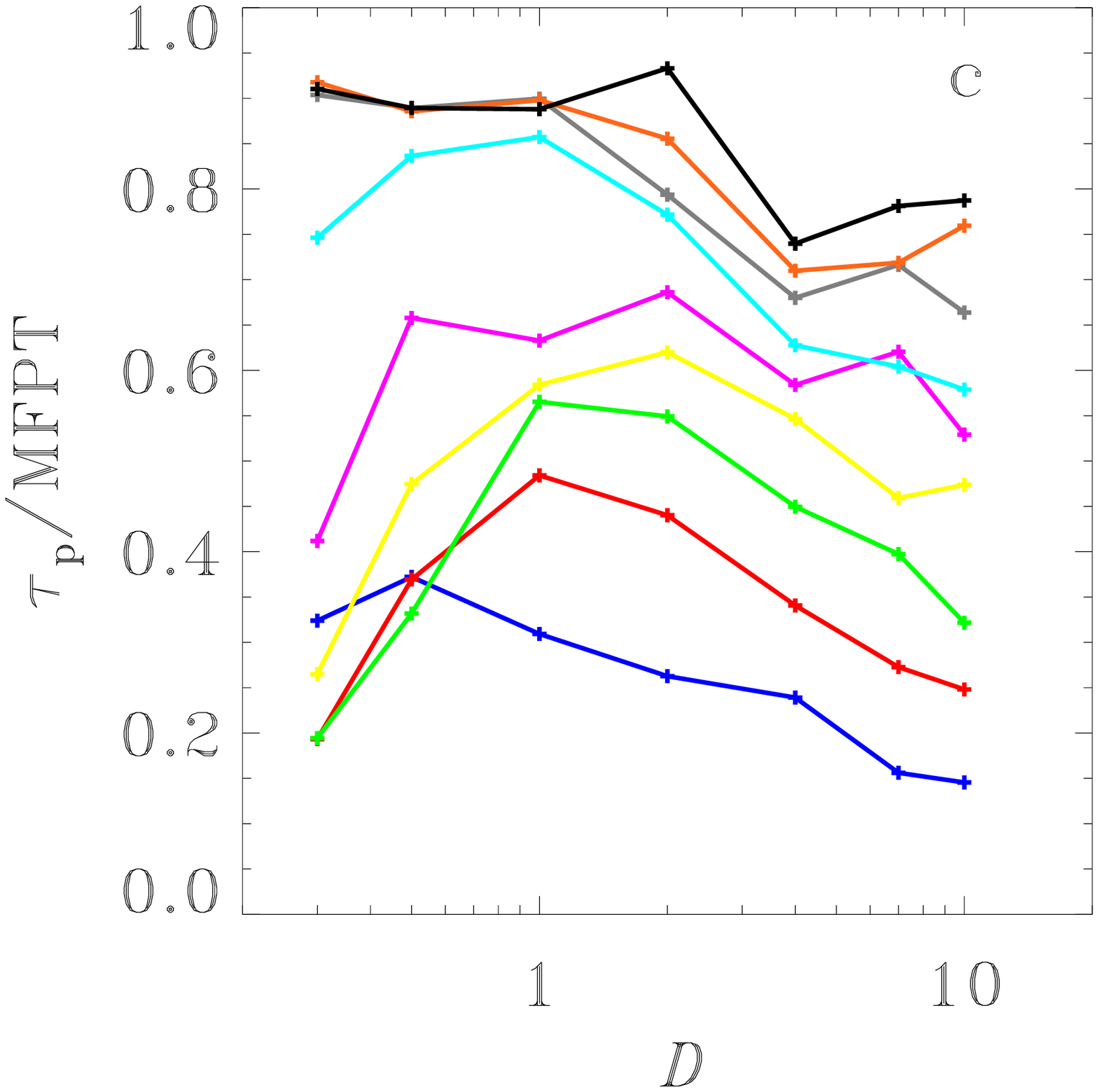}
\caption{\small Log-log plots of: a) Mean first passage time
vs.~noise intensity for 11 different values of the polymer length;
b) most probable translocation time vs.~noise intensity; c) ratio of
the most probable translocation time on MFPT vs.~noise intensity.
The colors correspond to different numbers $N$ of chain beads: blue
($N=4$), red ($N=6$), green ($N=8$), yellow ($N=10$), purple
($N=12$), cyan ($N=15$), grey ($N=20$), orange ($N=25$), black
($N=30$). } \label{fig2}
\end{figure}

The polymer center of mass crosses the line at $x_f$ with
translocation times having a distribution that crucially depends on
the noise intensity $D$. The mean crossing time, the median and
$\tau_{\rm p}$ are analyzed as a function of $D$ for different
values of the polymer length. In particular, figure \ref{fig2}a
shows a decreasing trend of the MFPT with increasing the intensity
of the fluctuations affecting the monomers motion. This result is
expected from the standard theory of the transport dynamics of a
particle escaping from a metastable state \cite{Gardiner1993}.
However, when the noise intensity goes down $D\simeq1$, the polymer
dynamics appears to be very sensitive to the molecule length, with
shorter chains ($N\lesssim12$) translocating more slowly (higher
MFPTs) compared to longer molecules. In particular, the dynamics of
polymers with $N\leqslant10$ and $D\leqslant1$ shows interesting
details inside the common behaviour, while the dependence of the
MFPT on $D$ converges to an almost unique path for greater values of
$N$. The crossing dynamics of chains with $N=4$ (blue line in figure
\ref{fig2}a) resembles that of a single particle (see figure 2 in
Ref.~\cite{Fiascon2003} and figure 1 in Ref.~\cite{Fiascon2005}), by
showing a marked power law relationship between the MFPT and the
noise intensity (straight line in a log-log plot). The comparison of
the polymer behaviour with that of a single particle started outside
the potential minimum, as studied in Refs.~\cite{Fiascon2003} and
\cite{Fiascon2005}, is possible because the exact initial point
becomes irrelevant at strong noise intensities, where the power law
behaviour is observed.

Many experimental works report their results in terms of the most
probable translocation time instead of the MFPT. For this reason, we
have also investigated how the median and $\tau_{\rm p}$ depend on
the noise intensity. The dependence of the median on $D$ is very
similar to that of MFPT, with a little exception at $D=10$. On the
contrary, the diagram of $\tau_{\rm p}$ vs. $D$ in figure
\ref{fig2}b shows some diversities. The single-particle-like
behaviour is still observed for $N=4$. However, when the polymer
length increases, the index of the power law between $\tau_{\rm p}$
and $D$ (the slope of the curve) progressively drops. For
$N\geqslant12$ the translocation dynamics changes with respect to
shorter chains and the dependence of $\tau_{\rm p}$ from the noise
intensity becomes almost equal to that observed for the MFPT. In
figure \ref{fig2}c the ratio $\tau_{\rm p}$/MFPT is plotted as a
function of $D$. We can see that the two characteristic crossing
times are not related to each other by a simple proportionality law
for different polymer lengths and noise intensities. However, a
common nonmonotonic behaviour of this ratio as a function of the
noise intensity is observed.

\begin{figure}[htbp]
\includegraphics[width=14cm,height=14cm]{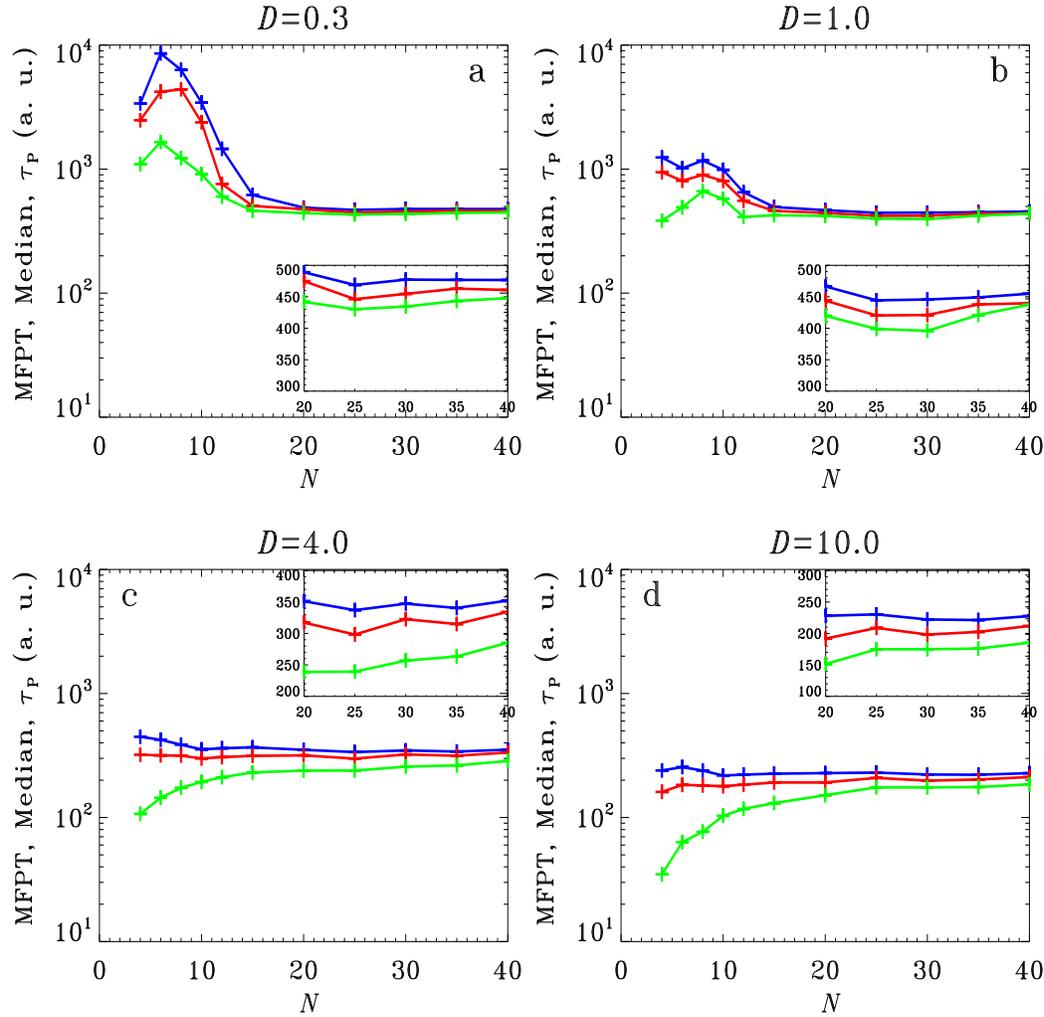}
\caption{\small Mean first passage time (blue line), median of the
crossing time distribution (red line) and most probable
translocation time (green line) vs.~polymer length (number of beads)
for four different noise intensity $D=0.3, 1.0, 4.0, 10.0$. The
inset on each diagram shows an enlarged view on the y-scale of the
region with $N>20$.} \label{fig3}
\end{figure}

The different dynamics of translocation of short and long polymers
is investigated as a function of the noise intensity. Figure
\ref{fig3} shows the dependence of the MFPT, the median and
$\tau_{\rm p}$ on the polymer length (the number $N$ of chain beads)
for four different values of noise intensity. In the diagram with
$D=0.3$ (figure \ref{fig3}a) two different regimes of translocation
are clearly present: for $N\lesssim15$ all characteristic times show
a maximum at $N\thickapprox7$, while for $N\gtrsim15$ a common
plateau is observed. This maximum is more evident in the MFPT than
in $\tau_{\rm p}$, but the height of both peaks is rapidly reduced
when the noise intensity increases. This maximum occurs at a polymer
length corresponding to the persistence length of the chain $L_{\rm
p}$, which represents the length over which the correlations in the
direction of the tangent to the polymer are lost. For our discrete
chain model, we have calculated the directional correlation:
\begin{eqnarray}
\langle \cos(\theta(s_j))\rangle=\frac{\sum_{k=0}^{N-j-1}
(x_k-\bar{x})(x_{k+j}-\bar{x})}{\sum_{k=0}^{N-1} (x_k-\bar{x})^2}
\label{eq8}
\end{eqnarray}
as a function of the contour length $s_j$ corresponding to the j-th
segment of the polymer chain, where $x_k$ represents
$\cos(\theta(s_k))$, $\theta(s_k)$ is the angle between the tangent
vector at one end of the chain molecule and the tangent vector at
the contour length corresponding to the k-th segment of the polymer
chain, $\bar{x}$ the mean of $x$. We have computed an ensemble
average over $5\times10^5$ different steric configurations of the
polymer fluctuating in a flat potential domain at the noise
intensity $D=0.3$ (the value for which the maximum of translocation
time is observed in figure \ref{fig3}a). The directional correlation
presents damped oscillations as a function of the polymer contour
length, following an overall exponentially-decreasing trend. By best
fitting this calculated correlation function with the expression
$\langle \cos(\theta(s))\rangle=exp(-s/L_{\rm P})$, which is valid
for the continuous worm-like chain model, we have obtained $L_{\rm
p}=29.8$. This value of $L_{\rm p}$ corresponds to $N\simeq6$.

In the short-length domain, the most probable translocation time
shows an inversion of the dependence from the polymer length at
higher values of $D$ with respect to that observed at lower noise
intensities. For $D\gtrsim2$ the most probable translocation time
increases monotonically with the chain length and, for $N\gtrsim15$,
it reaches almost the same level of the MFPT. The inset of each
diagram in figure \ref{fig3} shows the flat region of the
corresponding plot in the range $20 \leqslant N \leqslant 40$ in an
enlarged y-scaling. For any values of $D$, we observe a slightly
increasing trend of the characteristic crossing times for longer
molecules.

\section{Conclusions}\label{sect4}
In this paper we focus our study on the noise influence on the
translocation dynamics of short polymers. We model the molecule as a
chain of spatially-extended interacting beads. The transport
dynamics is simulated in a noisy environment and in the presence of
a potential barrier by solving the Langevin equation of motion for
every single monomer.

We find a maximum of the MFPT and $\tau_{\rm p}$ as a function of
the number of beads of the chain at low noise intensity. This
maximum occurs at a polymer length corresponding to the persistence
length of the chain $L_{\rm p}$. This characteristic length
represents a transition point between two different dynamical
behaviours of the polymer approaching the barrier. The chain
molecule having length lower than $L_{\rm p}$ moves like a flexible
elastic rod and the addition of a monomer to the chain causes longer
crossing times. For lengths greater than $L_{\rm p}$ the polymer can
fold, assuming a snake-like shape. This configuration helps the
molecule into the translocation process, reducing the crossing
times. For short molecules in little noisy environment, the
relationship between $\tau_{\rm p}$ and the chain length is similar
to that of MFPT. This means that both $\tau_{\rm p}$ and MFPT can be
used without distinction to investigate the polymer translocation
dynamics. On the contrary, when the thermal fluctuations are
stronger, the MFPT and $\tau_{\rm p}$ show two distinct dependencies
on $N$.

In order to compare our results with available experimental
findings, we restrict the following discussion on the behaviour of
the most probable time. In a low noise environment, we confirm that
short polymers can travel more slowly than longer ones, as firstly
observed by Han and collaborators (1999). For greater values of
polymer length ($N\gtrsim25$), our findings are in agreement with
the almost-linear relationship between the most probable
translocation time and the polymer length \cite{Kasianowicz1996}. At
higher temperatures, the fluctuations dominate the transport
dynamics of short polymers, canceling any dependence of the
translocation time on the stiffness of the molecule. In this regime,
longer molecules take longer times to cross the barrier.

For polymer chains having length in a range close to $L_{\rm p}$ the
noise acts as a trigger of the translocation dynamics. Low
intensities of thermal fluctuations bring the molecule to cross the
barrier by a transport mechanisms that is completely different from
that observed at higher temperatures. The measured crossing times
are a direct consequence of the noise influence on the molecule
dynamical regime of translocation. Further work is required to
include in our description a polymer-pore interaction term and to
explore the noise induced modifications of the polymer dynamics in a
sub-persistence length confinement.

The saturation of all the times represents an effect still not
completely understood. A possible explanation could be related to
the intrinsic stochastic behavior of the polymer dynamics: large
number of monomers gives an average contribution to the motion of
the polymer center of mass. When the number $N$ increases, the
fluctuations of the single monomer affect less and less the overall
motion of the polymer. This could be considered as a
"coarse-grained" one-particle description in the limit $N\to\infty$.
Moreover, the three characteristic translocation times saturates to
almost the same value, indicating that the escaping time probability
distributions tend to become more symmetric by increasing the number
of monomers. A detailed investigation of these features will be the
subject of a future work.

\ack This work was partially supported by MIUR and CNISM-INFM. N.~P.
wish to thank Dr.~Dominique Persano Adorno for helpful discussions.
A.~F. acknowledges the Marie Curie TOK grant under the COCOS project
(6th EU Framework Programme, contract No: MTKD-CT-2004-517186).


\section*{References}
\begin{thebibliography}{99}
\bibitem{Leng1998}
Lengauer C, Kinzler K W and Vogelstein B 1998 {\it Nature} {\bf
396}, 643

\bibitem{Sover2006}
Soverini S et al 2006 {\it Clin. Cancer Res.} {\bf 12}, 7374

\bibitem{Tseng2002}
Tseng Y, Liu J and Hong R 2002 {\it Mol. Pharmacol.} {\bf 62}, 864

\bibitem{Garcia2005}
Garcia-Garcia E, Andrieux K, Gil S, Couvreur P 2005 {\it Int. J.
Pharmaceutics} {\bf 298}, 274

\bibitem{Yotsu2008}
Yotsumoto S, Saegusa K and Aramaki Y 2008 {\it J. Immunol.}  {\bf
180}, 809

\bibitem{Higgins2007}
Higgins C F 2007 {\it Nature} {\bf 446}, 749

\bibitem{Sunda2008}
Sundaresan V B and Leo D J 2008 {\it Sens. Actuators B: Chem.} {\bf
131}, 384

\bibitem{Peyrin2001}
Peyrin E, Caron C, Garrel C, Ravel A, Villet A, Grosset C and Favier
A 2001 {\it Talanta} {\bf 55}, 291

\bibitem{Keller2007}
Keller S, B$\rm\ddot{o}$the M, Bienert M, Dathe M and Blume A 2007
{\it Chem. Bio. Chem.} {\bf 8}, 546

\bibitem{Mannion2006}
Mannion J T, Reccius C H, Cross J D and Craighead H G 2006 {\it
Biophys. J.} {\bf 90}, 4538

\bibitem{Kasianowicz1996}
Kasianowicz J J, Brandin E, Branton D and Deamer D W 1996 {\it Proc.
Natl. Acad. Sci.} USA {\bf 93}, 13770

\bibitem{Akeson1999}
Akeson M, Branton D, Kasianowicz J J, Brandin E and Deamer D W 1999
{\it Biophys. J.} {\bf 77}, 3327

\bibitem{Meller2000}
Meller A, Nivon L, Brandin E, Golovchenko J A and Branton D 2000
{\it Proc. Natl. Acad. Sci.} USA {\bf 97}, 1079

\bibitem{Meller2002}
Meller A and Branton D 2002 {\it Electrophoresis} {\bf 23}, 2583

\bibitem{Luo2008}
Luo K F, Ala-Nissila T, Ying S C and Bhattacharya A 2008 {\it Phys.
Rev. Lett.} {\bf 100}, 58101

\bibitem{Wanunu2008}
Wanunu M, Chakrabarti B, Math$\acute{e}$ J, Nelson D R and Meller A
2008 {\it Phys. Rev. E} {\bf 77}, 31904

\bibitem{Li2001}
Li J L, Stein D, McMullan C, Branton D, Aziz M J, Golovchenko J A
2001 {\it Nature} {\bf 412}, 166

\bibitem{Li2003}
Li J L, , Gershow M, Stein D, Brandin E, Golovchenko J A 2003 {\it
Nature Mater.} {\bf 2}, 611

\bibitem{Storm2005a}
Storm A J, Storm C, Chen J, Zandbergen H, Joanny J and Dekker C 2005
{\it Nano Lett.} {\bf 5}, 1193

\bibitem{Storm2005b}
Storm A J, Chen J, Zandbergen H and Dekker C 2005 {\it Phys. Rev. E}
{\bf 71}, 51903

\bibitem{Aksi2004}
Aksimentiev A, Heng J B, Timp G and Schulten K 2004 {\it Biophys.
J.} {\bf 87}, 2086

\bibitem{Fan2005}
Fan R, Karnik R, Yue M, Li D, Majumdar A and Yang P 2005  {\it Nano
Lett.} {\bf 5}, 1633

\bibitem{Folo2005}
Fologea D, Uplinger J, Thomas B, McNabb D S and Li J 2005 {\it Nano
Lett.} {\bf 5}, 1734

\bibitem{Han1999}
Han J, Turner S W and Craighead H G 1999 {\it Phys. Rev. Lett.} {\bf
83}, 1688

\bibitem{Han2000}
Han J and Craighead H G 2000 {\it Science} {\bf 288}, 1026

\bibitem{Sung1996}
Sung W and Park P J 1996 {\it Phys. Rev. Lett.} {\bf 77}, 783

\bibitem{Muthu1999}
Muthukumar M 1999 {\it J. Chem. Phys.} {\bf 111}, 10371

\bibitem{Lubensky1999}
Lubensky D K and Nelson D R 1999 {\it Biophys. J.} {\bf 77}, 99005

\bibitem{Tian2003}
Tian P and Smith G D 2003 {\it J. Chem. Phys.} {\bf 119}, 11475

\bibitem{Luo2006}
Luo K F, Ala-Nissila T and Ying S C 2006 {\it J. Chem. Phys.} {\bf
124}, 34714

\bibitem{Huop2007}
Huopaniemi I, Luo K F, Ala-Nissila T and Ying S C 2007 {\it Phys.
Rev. E} {\bf 75}, 61912

\bibitem{Luo2007}
Luo K F, Ala-Nissila T, Ying S C and Bhattacharya A 2007 {\it Phys.
Rev. Lett.} {\bf 99}, 148102

\bibitem{Park1998}
Park P J and Sung W 1998 {\it J. Chem. Phys.} {\bf 108}, 3013

\bibitem{Sebastian2006}
Sebastian L K and Debnath A 2006 {\it J. Phys.: Condens. Matter}
{\bf 18}, S283

\bibitem{Pizz2008}
Pizzolato N, Fiasconaro A, Spagnolo B 2008 Noise effects in polymer
dynamics, {\it Int. J. Bifurc. Chaos} Vol 18, in press

\bibitem{Meller2001}
Meller A, Nivon L and Branton D 2001 {\it Phys. Rev. Lett.} {\bf
86}, 3435

\bibitem{Rouse1953}
Rouse P E J 1953 {\it J. Chem. Phys.} {\bf 21}, 1272

\bibitem{Tothova2005}
Tothova J, Brotovsky B and Lisy V 2005 {\it Czech. J. Phys.} {\bf
55}, 221

\bibitem{Gardiner1993}
Gardiner C W 1993 {\it Handbook of stochastic methods for physics,
chemistry and the natural sciences} (Berlin) Springer.

\bibitem{Fiascon2003}
Fiasconaro A, Valenti D, Spagnolo B 2003 {\it Physica A} {\bf 325},
136

\bibitem{Fiascon2005}
Fiasconaro A, Spagnolo B, Boccaletti S 2005 {\it Phys. Rev. E} {\bf
72}, 061110

\end{thebibliography}
\end{document}